\documentclass[lettersize,journal]{IEEEtran}
\IEEEoverridecommandlockouts%

\usepackage{cite}
\usepackage{amsmath,amssymb,amsfonts}
\usepackage{algorithmic}
\usepackage{graphicx}
\usepackage{textcomp}
\usepackage{paralist}
\usepackage[dvipsnames, svgnames]{xcolor}
\usepackage[most]{tcolorbox}
\tcbuselibrary{breakable}

\tcbset{
  base/.style={
    arc=2mm,
    bottomtitle=0.3mm,
    boxrule=0mm,
    colbacktitle=black!10!white,
    coltitle=black,
    fonttitle=\bfseries,
    left=0.5mm, 
    lefttitle=0.5mm,
    leftrule=1mm,
    right=0.5mm, 
    righttitle=0.5mm,
    top=0.2mm,
    bottom=0.2mm,
    title={#1},
    toptitle=0.3mm,
  }
}

\tcbset{
  base_inline/.style={
    on line,
    arc=2mm,
    boxrule=0pt,
    leftrule=1mm,
    fontupper=\bfseries,
    boxsep=0pt,
    left=1.5mm,
    right=1.5mm,
    top=0.5mm,
    bottom=0.5mm,
  }
}
\newtcolorbox{rqbox}[1]{
  breakable,
  colframe=MidnightBlue,
  base={#1}
}
\newtcbox{\rqboxinline}{
    colframe=MidnightBlue,
    base_inline
}

\newcommand{\hquad}{\hspace{0.5em}}

\usepackage{fontawesome5}
\newcommand{\email}[1]{\href{mailto:#1}{\textcolor{black}{\texttt{#1}}}}

\newcommand{\orcidlink}[1]{\href{https://orcid.org/#1}{\textcolor[HTML]{A6CE39}{\faOrcid}}}
\newcommand{\mylink}[2]{\href{#1}{\textcolor{black}{\texttt{#2}}}}

\usepackage{subcaption}

\usepackage[T1]{fontenc}
\usepackage{inconsolata}

\usepackage[hidelinks]{hyperref}
\usepackage[capitalize]{cleveref} 

\begin{document}

\title{A Multi-Dimensional, Per-Pass Empirical Study of the LLVM Optimization Pipeline 
}

\author{Federico Bruzzone~\orcidlink{0009-0004-6086-8810} and Walter Cazzola~\orcidlink{0000-0002-4652-8113}
\thanks{F.~Bruzzone and W.~Cazzola are with the Computer Science Department, Universit\`a degli Studi di Milano, Milan, Italy (e-mail: \email{federico.bruzzone@unimi.it}; \email{cazzola@di.unimi.it}).}}

\markboth{Preprint, under review}%
{Bruzzone and Cazzola: A Multi-Dimensional, Per-Pass Empirical Study of the LLVM Optimization Pipeline}

\maketitle

\begin{abstract}
Quantifying the marginal impact of individual optimization passes underpins phase ordering, pass selection, optimization design, and analysis of pass/hardware interactions.
In LLVM, the standard backend for C/C++, Rust, and ML stacks via MLIR, interactions among passes, measurement noise, and pipeline scale make this difficult.
We present a systematic empirical study of the LLVM \texttt{-O3} pipeline. We decompose it into 113 cumulative per-pass prefixes and measure execution time, compile time, binary size, hardware counters, and RAPL energy on 30 PolyBench/C kernels (84,750 measurements).
Transitions whose binary is byte-identical to their predecessor give us a built-in null distribution, which we use to calibrate every significance claim.
On these compute-bound affine kernels, the pipeline is non-monotone: 36\% of the transitions that significantly change runtime are regressions, found in 28 of 30 kernels.
Gains are back-loaded (the median kernel needs 84.8\% of the pipeline for 80\% of its speedup) and concentrated in a small core of passes, a result that holds under both absolute and scale-free impact measures.
IR instruction count is an unreliable runtime predictor, and runtime-targeted passes are \textit{de facto} energy-targeted (30--60\% savings).
Finally, we bound the speedup lost to non-monotone behavior from above by 46.35\% and decompose it: 9.7 points are measurement noise, 15.3 points are transient dips repaired within three passes, and only 21.3 points persist to the end of the pipeline.
These findings inform pass pruning, cost-model calibration, and autotuning.
\end{abstract}


\section{Introduction}\label{sec:intro}
\IEEEPARstart{S}{ince} the dawn of computing, compiler optimizations have bridged high-level abstractions and efficient machine code~\cite{Ershov58, Heller61, McKeeman65, Gear65, Lowry69, Busam69}.
Optimizing compilers~\cite{Allen69, Wulf73b} use analysis passes to guide transformations~\cite{Cooper22, Kennedy01b, Paige96} that enhance performance without altering observable behavior~\cite{Bacon94, Dodds18, Aho06}, and modern architectures make this a delicate interplay of interdependent passes~\cite{Cooper22}.
LLVM~\cite{Lattner04} exemplifies this complexity. It is the standard backend for C/C++, Rust~\cite{Matsakis14}, Julia~\cite{Bezanson12}, and, via MLIR~\cite{Lattner21}, for \textit{machine learning} (ML) and \textit{deep learning} (DL) stacks~\cite{Paszke19, Abadi16} built on ML/DL optimizing compilers~\cite{Li21c, Xing19, Zhao18, Liu22d} such as OpenXLA\footnote{
  \mylink{https://openxla.org/}{openxla.org}
} and TVM~\cite{Chen18c}.

\noindent\textbf{The Phase-Ordering Problem.}\hquad%
The optimal sequence of optimizations is program-specific~\cite{Cooper99, Cooper02, Triantafyllis03}.
The \textit{phase-ordering problem}~\cite{Touati06, Freudenberger91, Zhao25}, i.e., finding the optimal pass order, is undecidable in general~\cite{Ashouri17, Davidson07, Vegdahl82}, and optimization selection is itself hard~\cite{Nobre16, Agakov06, Whitfield90}.
\textit{Compiler autotuning}~\cite{Pan25, Zhu24, Ashouri18, Tiwari09, Pan08, Pan06, Kulkarni04b} searches this large space~\cite{Purini13, Kulkarni12, Kulkarni03}, but its effectiveness depends on accurately estimating pass impact, which is complicated by inter-pass dependencies, noisy measurements, and scale~\cite{Whitfield97}. Some transformations even degrade performance~\cite{Almagor04, Zhao06}.
Despite this, both industry and academia still rely on predefined, \textit{one-size-fits-all} pipelines (e.g., \texttt{-O3}) for efficiency and because users have limited knowledge of the optimization process~\cite{Liu24, Georgiou18, Tan20}.

\noindent\textbf{Motivation and Gap.}\hquad%
Quantifying the impact of individual passes matters to compiler engineers and end users~\cite{Popescu25, Demertzi11, Cavazos07}. It enables
\begin{inparaenum}[(i)]
    \item better phase ordering,
    \item better pass-selection heuristics,
    \item informed optimization design, and
    \item analysis of pass/hardware interactions~\cite{Davidson07}.
\end{inparaenum}
Yet the folklore about \texttt{-O3} (``most passes do little'', ``passes sometimes undo each other'') has rarely been measured on a production pipeline.
The closest studies measure execution time only, either on a curated subset of LLVM passes~\cite{DeLaTorre18} or on GCC/ICC~\cite{Chen12}. They therefore cannot see trade-offs across objectives: a pass that improves runtime may increase binary size or degrade cache locality.
Phase-ordering work, in turn, quantifies loss only relative to an alternative order found by search~\cite{Kulkarni06, Almagor04}, which conflates the cost of the default order with the reach of the search.
Finally, pass-level studies rarely calibrate what counts as a ``real'' per-pass effect against measurement noise, so small deltas are easily over-interpreted.

\noindent\textbf{Our Study.}\hquad%
We address these gaps with a systematic empirical study of the full LLVM \texttt{-O3} pipeline (113 pass positions, not a curated subset), quantifying the effect of each pass on runtime, compile time, binary size, nine hardware counters, and RAPL energy\footnote{
Running average power limit~\cite{VanKempen25, Lee24, Pereira17}.
}.
We formulate the following research questions (RQs):

\begin{rqbox}{}
    \textbf{RQ\textsubscript{1}.} \textit{What is the marginal impact of individual optimization passes within the LLVM \texttt{-O3} pipeline?}
\end{rqbox}
\begin{rqbox}{}
    \textbf{RQ\textsubscript{2}.} \textit{To what extent do optimization passes trade off execution speed, compilation overhead, and binary footprint?}
\end{rqbox}
\begin{rqbox}{}
    \textbf{RQ\textsubscript{3}.} \textit{How do optimization passes affect hardware-level behavior as reflected by performance counters?}
\end{rqbox}
\begin{rqbox}{}
    \textbf{RQ\textsubscript{4}.} \textit{How much speedup does non-monotone pipeline behavior forfeit, how much of it persists to the end of the pipeline, and which passes are responsible?}
\end{rqbox}

\noindent To answer them, we built an infrastructure that decomposes an optimization pipeline into its constituent passes and builds a sequence of cumulative prefixes, each extending the previous one by a single pass. For every prefix, we measure the newly added pass relative to both the preceding prefix and the \texttt{-O0} baseline, enabling both a \textit{differential} per-pass analysis and a \textit{cumulative} analysis of the whole trajectory.
Because many passes leave the binary untouched on a given kernel, the study contains its own null distribution: we use the 2,434 transitions with byte-identical binaries to calibrate every significance threshold we apply.

\vspace{0.1em}\noindent\textbf{Contributions.}\hquad%
The paper contributions are:
\begin{compactitem}
   \item the first per-pass characterization of a complete production \texttt{-O3} pipeline across five dimensions (runtime, compile time, binary size, hardware counters, energy), including, to our knowledge, the first per-pass energy profile of LLVM;
   \item a noise-calibrated analysis of pass impact, showing that non-monotonicity, the concentration of gains in a small core of passes, and their back-loading are robust to estimator choice and to absolute versus scale-free measures, and that IR instruction count is an unreliable runtime proxy;
   \item a search-free upper bound \(L\) on the speedup forfeited by non-monotone behavior, decomposed per benchmark into measurement noise, transient dips, and persistent loss, and attributed to individual passes; and
   \item a reusable, configuration-driven infrastructure for pipeline decomposition and multi-metric measurement, released with all raw data.
\end{compactitem}

\noindent The rest of the paper is organized as follows. \S\ref{sec:llvm-pipeline} provides background, \S\ref{sec:methodology} describes the methodology, \S\ref{sec:results} presents the results, \S\ref{sec:discussion} discusses their implications, \S\ref{sec:threats} addresses threats to validity, \S\ref{sec:related} reviews related work, and \S\ref{sec:conclusion} concludes.

\section{The LLVM Optimization Pipeline}\label{sec:llvm-pipeline}
LLVM is a modular SSA-based compilation framework~\cite{Rosen88, Alpern88, Cytron89}. Its \texttt{opt} tool applies sequences of analysis and transformation passes that constitute the optimization pipeline.
The \textit{new} pass manager\footnote{
    \mylink{https://llvm.org/docs/NewPassManager.html}{llvm.org/docs/NewPassManager.html}
} organizes the pipeline hierarchically through nested pass managers operating at different IR granularities: \texttt{\textbf{module}}, \texttt{\textbf{function}}, \texttt{\textbf{cgscc}}, \texttt{\textbf{loop}}, and \texttt{\textbf{loop-mssa}}. Passes may therefore appear multiple times within the pipeline at different levels of the hierarchy.
At the top level, the \texttt{\textbf{module}} manager coordinates module-wide optimizations (e.g., \texttt{called-value-propagation}~\cite{Callahan86, Wegman91}) and invokes nested managers. The remaining managers apply passes that operate respectively on \begin{inparaenum}[(i)]
\item individual functions (e.g., \texttt{mem2reg} which promotes stack allocations to virtual registers~\cite{Cytron91, Muchnick97}, and \texttt{sroa} which performs \textit{scalar replacement of aggregates}~\cite{Carr94, Lu97}),
\item strongly connected components of the call graph (e.g., interprocedural \texttt{inline}~\cite{Chang89}),
\item loops (e.g., \texttt{loop-unroll}~\cite{Dongarra79, Davidson95}), and
\item loops in memory SSA form (e.g., \texttt{licm} which performs \textit{loop-invariant code motion}~\cite{Knoop94, Allen72}).
\end{inparaenum}
Passes execute sequentially over an evolving IR\@. Consequently, the effect of a pass depends on the transformations that precede it, making optimization pipelines inherently \textit{order-dependent}, \textit{non-commutative}, and \textit{non-additive}.

\section{Experimental Methodology}\label{sec:methodology}
We develop an infrastructure that decomposes LLVM pipelines into passes and measures their marginal impacts.
The four RQs introduced in~\S\ref{sec:intro} are operationalized as follows.

\noindent\rqboxinline{RQ\textsubscript{1}.} We quantify the marginal impact of each optimization pass by measuring the delta of its performance relative to both the preceding pipeline state and the \texttt{-O0} baseline.

\noindent\rqboxinline{RQ\textsubscript{2}.} We quantify pass-level trade-offs across execution time, compile time, and binary size.

\noindent\rqboxinline{RQ\textsubscript{3}.} We characterize the effect of optimization passes on hardware behavior by tracking the evolution of performance counters across the pipeline.

\noindent\rqboxinline{RQ\textsubscript{4}.} We bound the speedup forfeited by non-monotone behavior with a search-free metric \(L\), decompose it into noise, transient, and persistent components, and attribute it to pass positions.

\noindent\textbf{Evaluation Infrastructure.}\hquad%
An \textit{orchestrator} coordinates pipeline construction, execution, and metric collection. Benchmarks, metrics, and experimental parameters are specified in TOML, allowing new studies without modifying the core implementation.
Although pipeline-agnostic, in this study the infrastructure targets LLVM's \texttt{-O3} pipeline. It first extracts and flattens the pipeline into an ordered sequence of passes \(p_0, p_1, \ldots, p_n\), then constructs cumulative prefixes \(P_i = p_0 \rightarrow \cdots \rightarrow p_i\), preserving the original execution order and pass dependencies.
The design is similar to that of Chen \textit{et al.}~\cite{Chen12}, who measured execution time only on GCC/ICC. We extend it to the complete LLVM pipeline and to five dimensions.
Each prefix is applied independently to the same unoptimized IR \(I_0\), producing an optimized IR \(I_i\) and executable binary \(B_i\). Comparing consecutive binaries \(B_i\) and \(B_{i-1}\) yields the marginal impact of pass \(p_i\) within its execution context.
We chose cumulative prefixes over leave-one-out ablation because they measure each pass exactly as \texttt{-O3} deploys it, after every pass that precedes it, and because the resulting trajectory is what RQ\textsubscript{4} requires. Leave-one-out answers a different, counterfactual question (what if \(p_i\) were deleted?), which mixes the effect of \(p_i\) with the reaction of all later passes to its absence.
Each flattened entry keeps its pass-manager nesting (e.g., \texttt{function(\ldots)}), so a function pass in prefix \(P_i\) runs on every function, exactly as in \texttt{-O3}. Analyses are computed on demand by the analysis manager and are not enumerated as separate entries.

\noindent\textbf{Benchmark Selection.}\hquad%
We use the 30 kernels of PolyBench/C 4.2.1\footnote{
    \mylink{https://github.com/MatthiasJReisinger/PolyBenchC-4.2.1/blob/master/polybench.pdf}{github.com/MatthiasJReisinger/PolyBenchC-4.2.1/polybench.pdf}
} in default mode. To keep the 95\% confidence interval below 1\% of the mean, benchmarks must be self-contained (single translation unit, \texttt{libm} only), deterministic (single-threaded), and execution-time tractable. PolyBench satisfies these requirements, whereas larger suites (e.g., SPEC or MultiSource) would either mask single-pass effects or exceed our repetition budget.
Despite its modest size, PolyBench covers a heterogeneous set of workloads~\cite{Yuki14}. Its six categories span the computation patterns most targeted by \texttt{-O3}, including loop nests, affine accesses, reductions, stencils, and triangular solvers. PolyBench is also the \textit{de facto} benchmark suite for autotuning and ML-based pass-selection studies~\cite{Ashouri18}, facilitating comparison with prior work.
We use the default dataset configuration, which maximizes level-1 cache (L1) utilization while keeping working sets within the last-level cache (LLC), thereby exposing optimization effects without being dominated by memory bottlenecks.
PolyBench is also relevant to the MLIR-based argument of~\S\ref{sec:intro}. Modern ML stacks lower tensor operations to LLVM IR through MLIR, where code generation is ultimately driven by the same \texttt{-O3} pipeline studied here. Many MLIR-generated kernels are structurally similar to PolyBench kernels: matrix multiplications and convolutions map naturally to \texttt{gemm}, \texttt{2mm}, and \texttt{3mm}; attention and reduction patterns resemble \texttt{atax}, \texttt{bicg}, \texttt{gesummv}, and \texttt{mvt}; normalization workloads resemble \texttt{correlation} and \texttt{covariance}; and iterative PDE-style computations resemble \texttt{jacobi-1d}, \texttt{jacobi-2d}, \texttt{heat-3d}, \texttt{seidel-2d}, and \texttt{fdtd-2d}. These structural correspondences suggest that our findings may transfer to MLIR-based ML pipelines, although direct validation remains future work.

\noindent\textbf{Metric Collection.}\hquad%
For each pipeline variant, we measure execution time, binary size, and end-to-end compilation time (wall-clock time for \texttt{clang\,+\,opt\,+\,llc}, excluding linking).
Hardware behavior is characterized using \texttt{perf stat} and nine validated events~\cite{Weaver08}: cycles and retired instructions, whose ratio gives \textit{instructions per cycle} (IPC, a throughput measure) and its inverse CPI; L1 data-cache (D1), L1 instruction-cache (L1-I), and last-level cache (LLC) misses, which expose data and code locality; branch misses, which expose control-flow predictability; and package energy via RAPL (\texttt{power/energy-pkg/}, DRAM/GPU excluded).
Runtime is roughly cycles divided by frequency, so a pass can speed up a kernel either by removing instructions or by raising IPC; the counters tell the two apart.
Cache references, peak RSS, page faults, and context switches are collected as confound checks.

\noindent\textbf{Experimental Setup.}\hquad%
All experiments ran on Intel Core i9-12900KF (Alder Lake, 8P+8E cores, 5.2\,GHz max), 128\,GB RAM, Fedora Linux 43 (kernel 6.19), with cache hierarchy: 640\,KiB D1, 768\,KiB L1-I, 14\,MiB L2, 30\,MiB L3. 
We used LLVM 21.1.8 (release, no assertions, no PGO). The \texttt{-O3} pass sequence was extracted via \texttt{opt --print-pipeline-passes -O3} and flattened.
The pipeline has 113 distinct invocations (including repeats of \texttt{instcombine}, \texttt{simplifycfg}, \texttt{early-cse}).
Our protocol entails \(84{,}750\) measurements (\(50{,}850\) runtime, \(33{,}900\) hardware-counter runs).
Runtime measurements use \texttt{hyperfine} (3 warmup, 15 timed runs), while hardware counters are collected with \texttt{perf stat} (2 warmup, 10 timed runs) in a separate batch. Ten runs are sufficient to keep the 95\% confidence interval below 1\% of the mean for all counter metrics.

\noindent\textbf{Noise Mitigation.}\hquad%
To reduce measurement variability, benchmarks are pinned to a dedicated P-core using \texttt{taskset}, executed under real-time FIFO scheduling (\texttt{chrt}), run with the CPU governor set to \texttt{performance} (\texttt{cpupower}), and evaluated with address-space layout randomization (ASLR) disabled (\texttt{setarch -R}). Hardware performance counters are collected in separate runs to avoid instrumentation overhead.

\noindent\textbf{Analysis Protocol and Null Calibration.}\hquad%
The 30 kernels and 113 prefixes yield 3,390 \textit{transitions} (\texttt{-O0} to prefix~0, and each prefix to the next).
For transition \(i\) we use the scale-free relative delta \(r_i = (T_i - T_{i-1})/T_{i-1}\) on median runtime, so \(r_i>0\) is a slowdown.
A transition is a \textit{significant} change when \(|r_i|>1\%\) and a two-sided Mann--Whitney \(U\) test on the 15 timed runs gives \(p<0.05\). Benjamini--Hochberg correction across all transitions changes the counts by at most one.
We calibrate this criterion on a natural null: 2,434 transitions (71.8\%) produce a binary byte-identical to the previous one, so any measured difference is noise.
On these, \(p<0.05\) alone fires in 25.7\% of cases, reflecting slow drift between measurement batches~\cite{Mytkowicz09}, while our combined criterion fires in only 22 cases (0.9\%). We therefore never rely on \(p\)-values or effect sizes alone, and report 0.9\% as the empirical false-positive rate of every count below.

Artifacts, including the enumerated pass sequences, are available on GitHub\footnote{\mylink{https://github.com/FedericoBruzzone/llvm-passview}{github.com/FedericoBruzzone/llvm-passview}} and Zenodo\footnote{\mylink{https://doi.org/10.5281/zenodo.20799028}{https://doi.org/10.5281/zenodo.20799028}}.

\begin{figure}[t]
    \centering
    \begin{subfigure}{\columnwidth}
        \centering
        \includegraphics[width=.9\linewidth, trim={0.25cm 0.25cm 0.25cm 0.25cm}, clip]{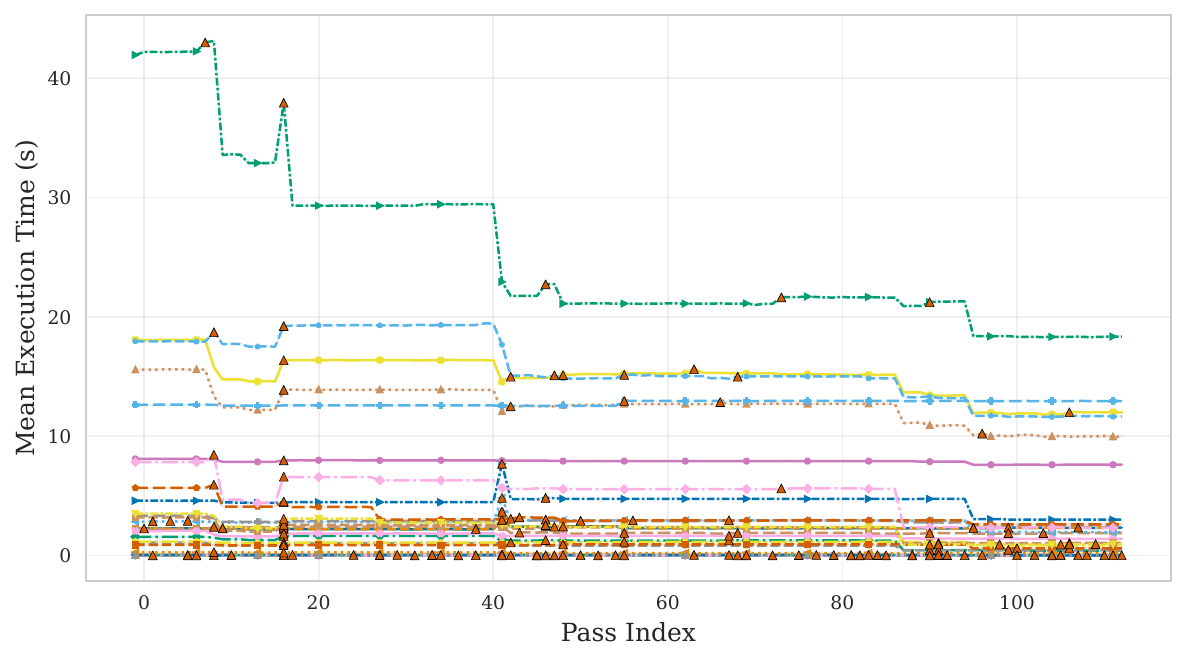}
        \caption{runtime evolution with 95\% confidence interval.}%
        \label{fig:evolution}
    \end{subfigure}

    \vspace{0.3cm}

    \begin{subfigure}{\columnwidth}
        \centering
        \includegraphics[width=.9\linewidth, trim={0.25cm 0.25cm 0.25cm 0.25cm}, clip]{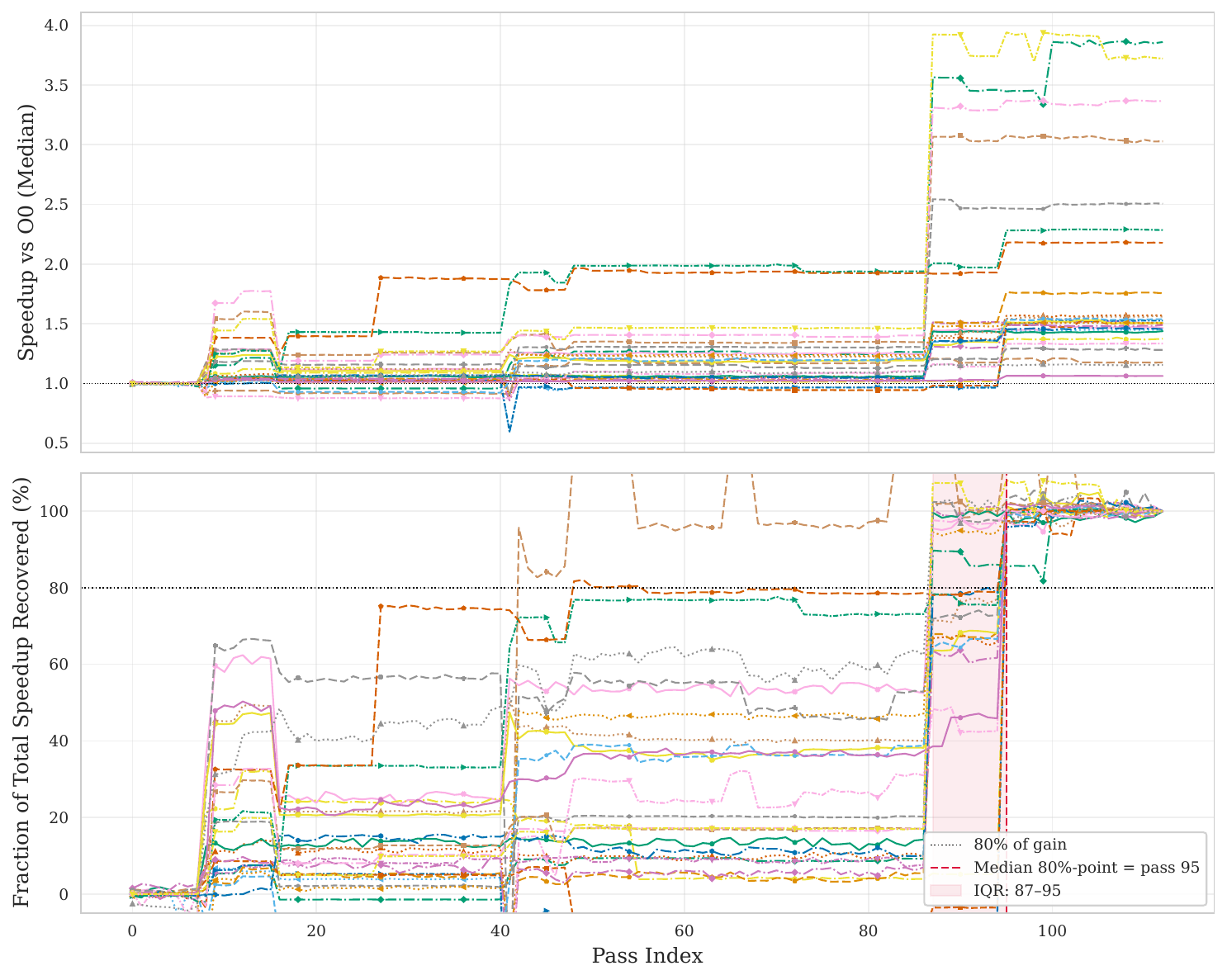}
        \caption{cumulative speedup \(S_i\) (\textit{top}) and normalized gain \(G_b(i)\) (\textit{bottom}).}%
        \label{fig:cumulative_speedup}
    \end{subfigure}

    \caption{Runtime evolution and optimization saturation.}%
    \label{fig:performance_analysis}
\end{figure}

\section{Results}\label{sec:results}
We report median statistics as recommended in~\cite{Georges07, Mytkowicz09}, and mean statistics with a 95\% confidence interval.
All multi-benchmark figures use the following legend (unique color, marker, line style per kernel):\\
\includegraphics[width=\columnwidth, trim={0.25cm 1cm 0.25cm 0.5cm}, clip]{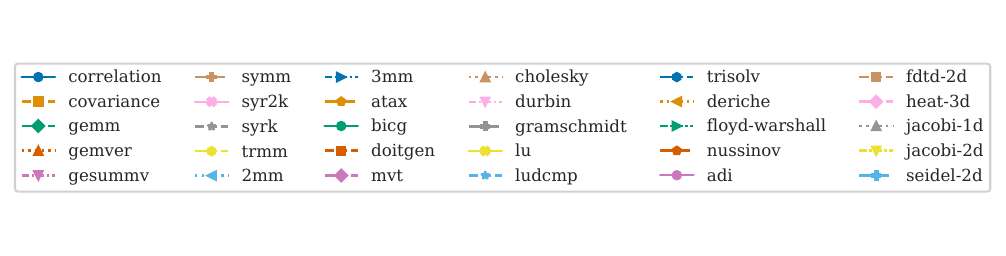}

\subsection{RQ\textsubscript{1}: Marginal Performance Impact}\label{sec:rq1}
\cref{fig:evolution} plots, for each kernel, median execution time (\(y\)) against the prefix index \(i\in[0,112]\) (\(x\)), with \(i=-1\) denoting \texttt{-O0}; shaded bands are 95\% confidence intervals, which stay below 1\% of the mean throughout.
\cref{fig:cumulative_speedup} shows the same trajectories as speedup over \texttt{-O0}, \(S_b^i\) (top), and as normalized gain \(G_b^i=(S_b^i-1)/(S_b^*-1)\in[0,1]\) (bottom), the fraction of the final speedup \(S_b^*\) reached at prefix~\(i\).

\noindent\textit{Non-monotonicity.}\hquad%
Of the 3,390 transitions, 291 significantly change runtime (\S\ref{sec:methodology}), and 106 of these (36\%) are \textit{regressions}.
Restricting to the 956 transitions that change the binary leaves 97 regressions and 172 improvements, again 36\%. Regressions occur in 28 of 30 kernels, have median magnitude \(+2.7\%\) (maximum \(+73\%\)), and are far above the 0.9\% false-positive rate of the null.
Three passes cause 68\% of them: \texttt{instcombine} (39), \texttt{simplifycfg} (16), and \texttt{licm} (11).
Counting every nominal increase (\(r_i>0\)) instead would label 48\% of all transitions as regressions, most of them noise.

\noindent\textit{Back-loading.}\hquad%
Across the 27 kernels whose final speedup exceeds~1, the median prefix needed to reach 80\% of the final speedup contains 84.8\% of the passes (IQR 77.7\%--84.8\%). Removing the last \({\sim}15\%\) of the pipeline thus sacrifices roughly 20\% of the attainable speedup, contradicting the intuition that most gains occur early.
The \(G_b\) curves have an S-shaped profile: an initial canonicalization phase, a steep growth region dominated by loop transformations and vectorization, and a final plateau. At \({\sim}25\%\) of the pipeline almost all kernels remain at \(S\approx1\); by \({\sim}50\%\), 70\% have crossed \(1.2\times\); at the end, 80\% reach \(1.5\times\) or more.
Several curves temporarily exceed \(G_b=1\): intermediate states that outperform the final binary, which we quantify in \S\ref{sec:rq2} and \S\ref{sec:rq4}.
Gains concentrate on dense-matrix and stencil kernels (\texttt{gemm} \(3.86\times\), \texttt{jacobi-2d} \(3.72\times\), \texttt{heat-3d} \(3.37\times\), \texttt{fdtd-2d} \(3.02\times\)), while \texttt{correlation} (\(0.96\times\)), \texttt{covariance}, and \texttt{seidel-2d} (\(0.98\times\)) finish slower than \texttt{-O0}.

\begin{figure}[t]
    \centering
    \begin{subfigure}{\columnwidth}
        \centering
        \includegraphics[width=\linewidth]{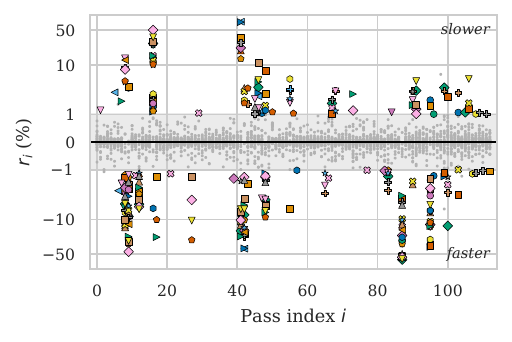}
        \caption{relative per-pass delta \(r_i\) for every transition (symmetric-log scale).}%
        \label{fig:waterfall}
    \end{subfigure}

    \vspace{0.2cm}

    \begin{subfigure}{\columnwidth}
        \centering
        \includegraphics[width=\linewidth]{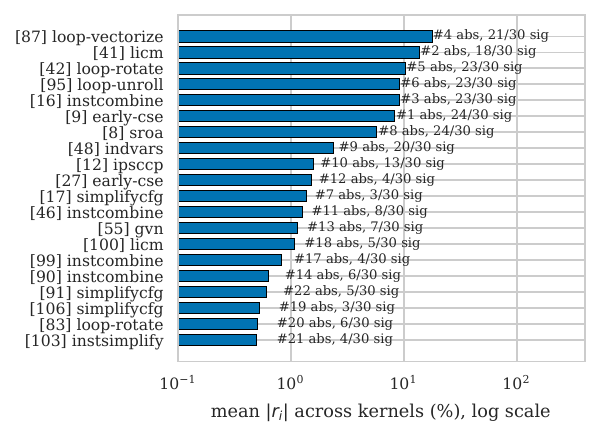}
        \caption{top-20 pass positions by mean \(|r_i|\), with rank under the absolute (ms) measure and number of kernels where the transition is significant.}%
        \label{fig:pass-ranking}
    \end{subfigure}
    \caption{Scale-free marginal impact of each pass.}
\end{figure}

\noindent\textit{Concentration.}\hquad%
\cref{fig:waterfall} plots \(r_i\) for every transition: grey dots are not significant (the shaded band is \(\pm1\%\)), colored markers are significant changes of the kernel given in the legend, and slowdowns lie above zero.
Most transitions are inert. Only 32.6\% modify the IR and 28.2\% the binary; 58 of the 112 pass positions never modify the IR of any kernel, and only 21 modify all 30.
Significant changes cluster in three bursts: early canonicalization (passes 8--18), loop canonicalization (41--48), and vectorization/unrolling (87--95).
\cref{fig:pass-ranking} ranks pass positions by mean \(|r_i|\) across kernels. Being relative, this measure weights every kernel equally regardless of its runtime scale.
It agrees with the absolute-ms ranking (Spearman \(\rho=0.79\); 9 of the top-10 and 18 of the top-20 positions coincide), so concentration is not an artifact of long-running kernels.
Under the relative measure, 14 positions (12.4\%) account for 80\% of the total impact, and within each kernel a median of 10.6\% of the passes (IQR 5.5\%--15.9\%) account for 80\% of \(\sum_i|\log(T_i/T_{i-1})|\).
The core is \texttt{loop-vectorize}, \texttt{licm}, \texttt{loop-rotate}, \texttt{loop-unroll}, \texttt{instcombine}, \texttt{early-cse}, and \texttt{sroa}. The two measures differ only in emphasis: \texttt{early-cse} ranks first in milliseconds because it fires on long-running kernels, whereas \texttt{loop-vectorize} ranks first in relative terms.
The column of significant kernels separates \textit{universal} passes (e.g., \texttt{early-cse}, \texttt{sroa}, 24/30) from \textit{program-specific} ones (e.g., \texttt{gvn}, 7/30).
The ranking is \textit{per occurrence}: passes scheduled several times (\texttt{instcombine}, \texttt{simplifycfg}, \texttt{licm}, \texttt{early-cse}) receive one entry per position, and only the first occurrence of each is in the core.
We do not report Cohen's \(d\): near-deterministic transitions have vanishing variance, which inflates \(|d|\) without bound, and the null shows that significance alone is unreliable.
Across categories, \texttt{blas} and \texttt{stencils} see large late-pipeline gains (medians above \(1.5\times\) and \(2.0\times\)), \texttt{datamining} stays near \(1.0\times\), and \texttt{solvers} and \texttt{medley} improve near-monotonically.

\begin{rqbox}{Finding 1 (RQ\textsubscript{1})}
The \texttt{-O3} pipeline is \textit{non-monotone}: 36\% of the transitions that significantly change runtime are regressions, present in 28 of 30 kernels and caused mostly by \texttt{instcombine}, \texttt{simplifycfg}, and \texttt{licm}.
Impact is concentrated: about 12\% of the pass positions account for 80\% of it, under both absolute and scale-free measures, while two thirds of the transitions do not even modify the IR.
The realized gain is \textit{back-loaded}: of the 27 kernels that speed up, 24 cross \(G_b{=}0.8\) right after \texttt{loop-vectorize} (pass~87, 10 kernels) or \texttt{loop-unroll} (pass~95, 14 kernels), placing the median crossing at 84.8\% of the pipeline.
\end{rqbox}

\begin{figure}[t]
    \centering
    \includegraphics[width=0.9\linewidth, trim={0.25cm 0.25cm 0.25cm 0.25cm}, clip]{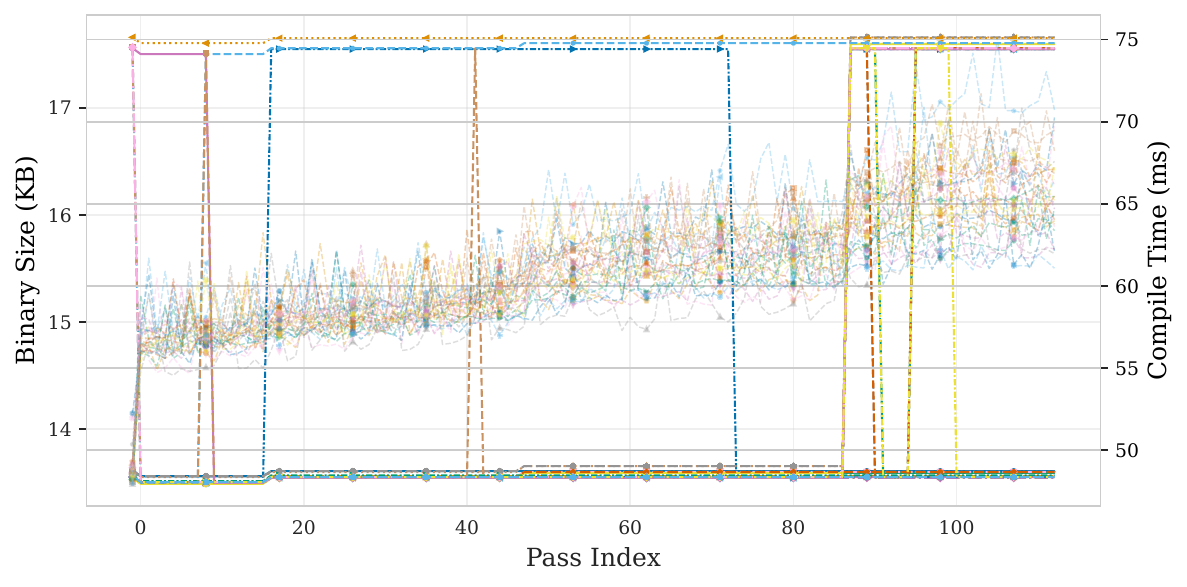}
    \caption{Compile time and binary size trajectories.}%
    \label{fig:tradeoff}
\end{figure}

\subsection{RQ\textsubscript{2}: Multi-Dimensional Trade-offs}\label{sec:rq2}
\begin{figure}[t]
    \centering
    \includegraphics[width=0.9\linewidth, trim={0.25cm 0.25cm 0.25cm 0.25cm}, clip]{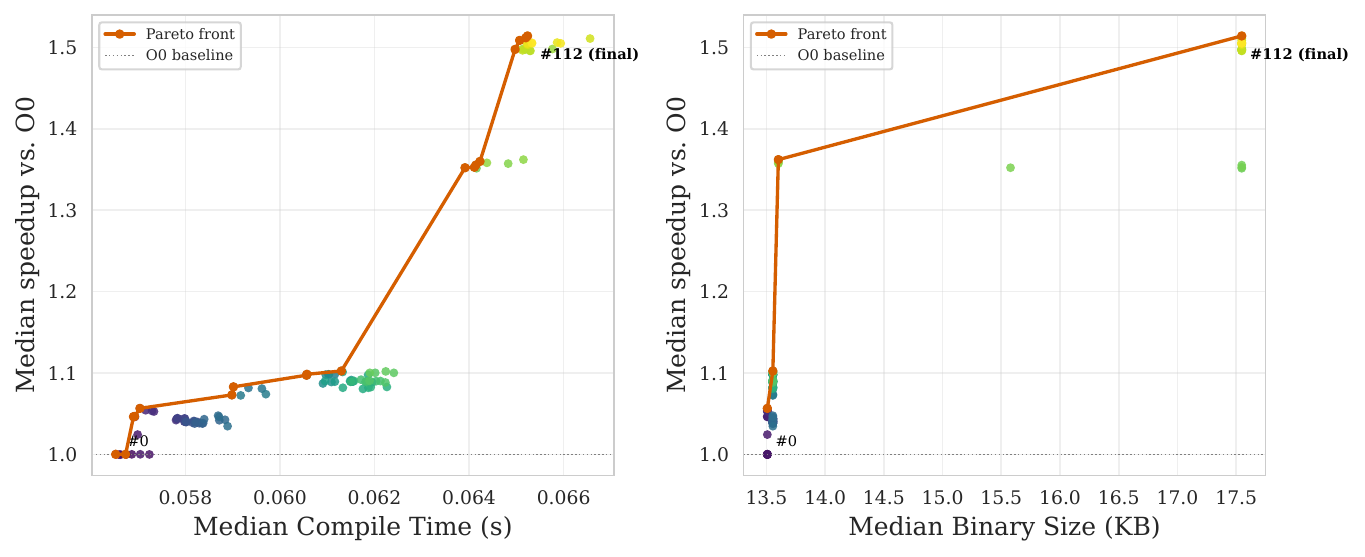}
    \caption{Pareto views of the multi-objective trade-off, each point aggregating all benchmarks at one variant index.}%
    \label{fig:pareto}
\end{figure}

RQ\textsubscript{2} asks whether runtime improvements are bought at the price of compile-time or footprint regressions.
In \cref{fig:tradeoff}, the bold step lines (left axis) are per-kernel binary sizes and the thin faded lines (right axis) are per-kernel compile times.
Compile time grows near-monotonically, accelerating where loop and inlining passes fire. Binary size is \textit{non-monotone}: early passes shrink it, while late vectorization and unrolling inflate it in a few discrete steps.
\cref{fig:pareto} aggregates all kernels per prefix (median speedup against median compile time, left, and median binary size, right; the orange line is the Pareto front).
The compile-time frontier is L-shaped: most speedup arrives at the pipeline tail (\({\sim}\)prefix~80) for modest extra compile cost. The size frontier is two-step: an initial wave of speedups is free in size, while the final jumps coincide with abrupt size increases.

\begin{figure}[t]
    \centering
    \includegraphics[width=0.9\linewidth, trim={0.25cm 0.25cm 0.25cm 0.25cm}, clip]{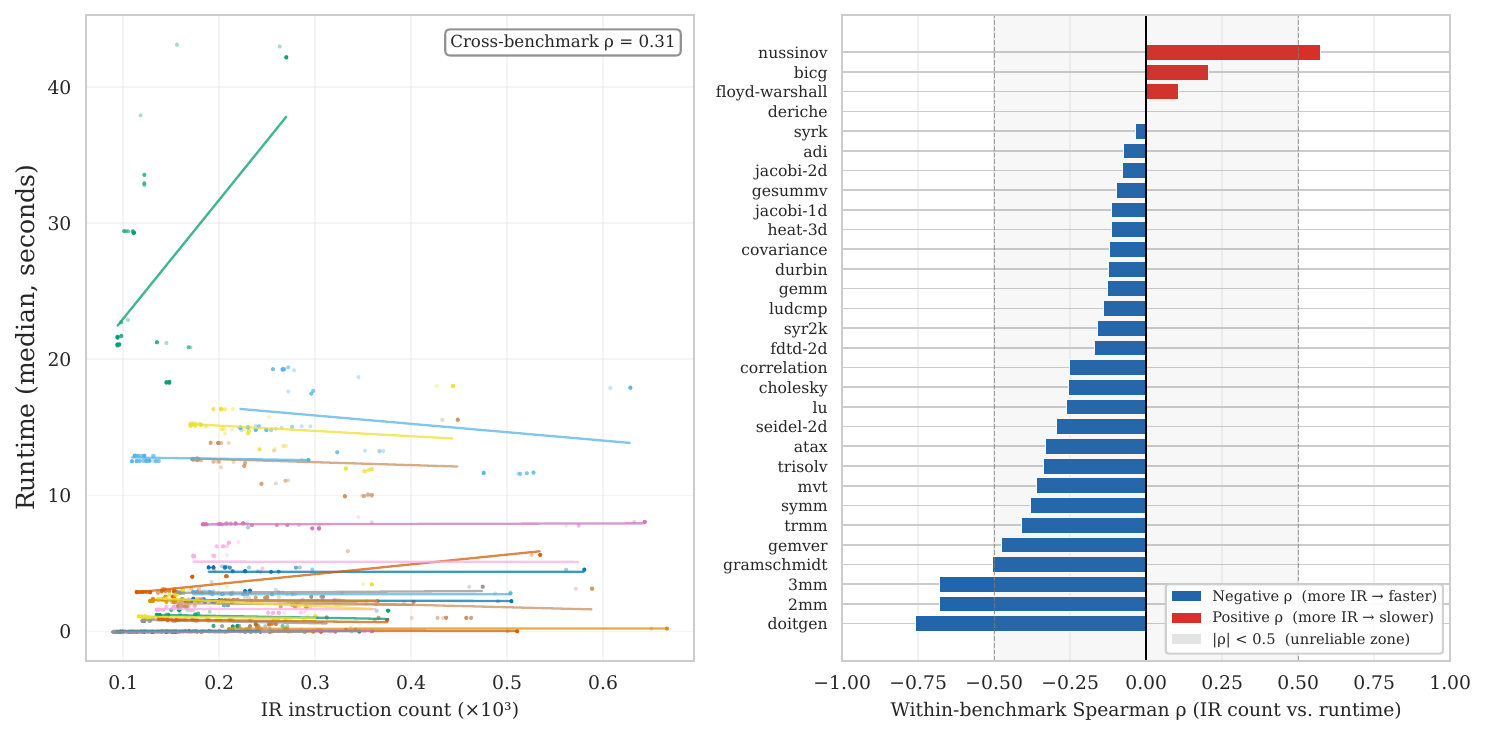}
    \caption{\textit{Left:} IR instruction vs.\ median runtime by benchmark; cross-benchmark Spearman \(\rho=0.31\). \textit{Right:} within-benchmark Spearman \(\rho\) for all 30 benchmarks.}%
    \label{fig:ir-vs-runtime}
\end{figure}

\noindent\textit{Is the final \texttt{-O3} binary Pareto-optimal?}\hquad%
Per kernel, in 29 of 30 cases some earlier prefix is nominally faster than the final binary at no larger size.
This dominance is real but thin. The best such prefix is significantly faster in 24 kernels, but only by a median 0.8\% (more than 1\% in 10 kernels), and it usually lies in the last 15 passes (median position 103), after binary size has reached its final value.
Only four kernels have a prefix that is strictly smaller \textit{and} faster: \texttt{correlation} (\(+6.5\%\) speed), \texttt{covariance} (\(+5.1\%\)), \texttt{seidel-2d} (\(+3.1\%\)), and \texttt{symm} (\(+2.5\%\)). They are the kernels with the largest persistent loss in \S\ref{sec:rq4}.

\noindent\textit{IR size as a runtime proxy.}\hquad%
\cref{fig:ir-vs-runtime} shows that static IR-instruction count cannot serve as a proxy for runtime.
The left panel (IR count against median runtime, one fitted trend per kernel) yields a weak cross-kernel Spearman \(\rho = 0.31\).
The right panel gives the within-kernel \(\rho\): 27 of 30 kernels have \textit{negative} \(\rho\) (more IR, faster code), because the passes that expand the IR (unrolling, vectorization) are those that reduce runtime. Only \texttt{nussinov}, \texttt{bicg}, and \texttt{floyd-warshall} are positive, and most bars fall within the \(|\rho|<0.5\) zone regardless of sign.

\begin{rqbox}{Finding 2 (RQ\textsubscript{2})}
Compile time grows near-monotonically, and binary size is non-monotone: early passes shrink it, late vectorization and unrolling inflate it.
The final \texttt{-O3} binary is nominally Pareto-dominated on (size, speedup) in 29 of 30 kernels, but the margin is significant in 24 and exceeds 1\% in only 10. A strictly smaller and faster prefix exists only for \texttt{correlation}, \texttt{covariance}, \texttt{seidel-2d}, and \texttt{symm} (2.5--6.5\% faster).
The \texttt{opt} stage accounts for less than half of total compile time, and its share is essentially kernel-independent, so pass-budget decisions can be made per pipeline, not per program.
Static IR-instruction count is \textit{not} a reliable proxy for runtime under pipeline-prefix variation.
\end{rqbox}

\begin{figure*}[tbh!]
    \centering
    \includegraphics[width=0.9\textwidth, trim={0.25cm 0.25cm 0.25cm 0.25cm}, clip]{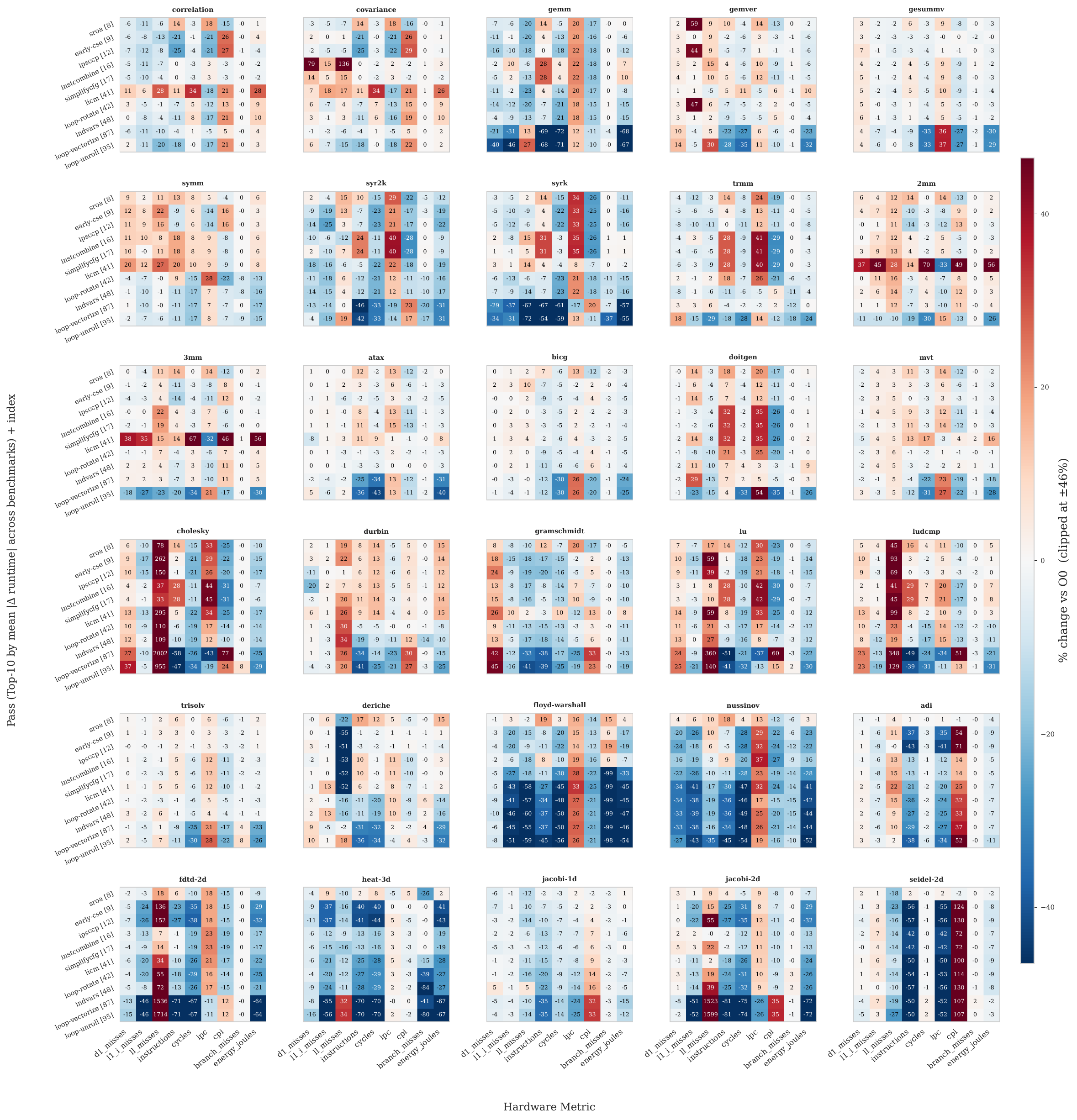}
    \caption{Per-benchmark \(5\!\times\!6\) grid of hardware-counter signatures at the top-10 most impactful pass positions.}%
    \label{fig:hw-heatmap}
\end{figure*}

\subsection{RQ\textsubscript{3}: Hardware-Level Insights}\label{sec:rq3}
RQ\textsubscript{3} asks how IR-level transformations propagate to hardware behavior.
\cref{fig:hw-heatmap} has one panel per kernel; rows are the ten pass positions with the largest absolute impact (nine of which are also in the relative top-10 of \cref{fig:pass-ranking}) and columns are counters. Each cell is the percentage change relative to \texttt{-O0}, blue for a decrease and red for an increase, with the colormap anchored at the 95th-percentile magnitude so that outliers do not flatten the remaining cells.
The top-10 positions form three groups.
Scalar canonicalization (8~\texttt{sroa}, 9~\texttt{early-cse}, 16~\texttt{instcombine}, 17~\texttt{simplifycfg}) gives the most uniform IPC gain (median \(+11\)--\(14\%\), 26--28/30 kernels) at the cost of transient instruction inflation (\(+5\)--\(13\%\)), while cycles still fall by 3--4\%.
Propagation (12~\texttt{ipsccp}) consistently reduces instructions (6--7\%), cycles (5\%), and energy (4--6\%) in 19--28 kernels.
Loop transformations and vectorization (41~\texttt{licm}, 42~\texttt{loop-rotate}, 48~\texttt{indvars}, 87~\texttt{loop-vectorize}, 95~\texttt{loop-unroll}) span the widest range. Pass~41 is the most variable (cycles \(-47\%\) to \(+70\%\)), while passes~42 and~48 are uniformly beneficial (instructions \(-7\) to \(-11\%\), energy \(-8\) to \(-9\%\), 25--28 kernels).
Pass~95 gives the deepest suite-wide reductions (median instructions \(-37\%\) in 29/30 kernels, cycles \(-33\%\), energy \(-29\%\)) but splits the LLC column: deep blue for most kernels, deep red for the \texttt{lu}/\texttt{ludcmp}/\texttt{cholesky} triad, where aggressive unrolling of dense triangular loops expands the data working set. The same pass thus produces qualitatively different microarchitectural signatures across kernels.

\begin{figure*}[t]
    \centering
    \includegraphics[width=0.9\textwidth, trim={0.25cm 0.25cm 0.25cm 0.25cm}, clip]{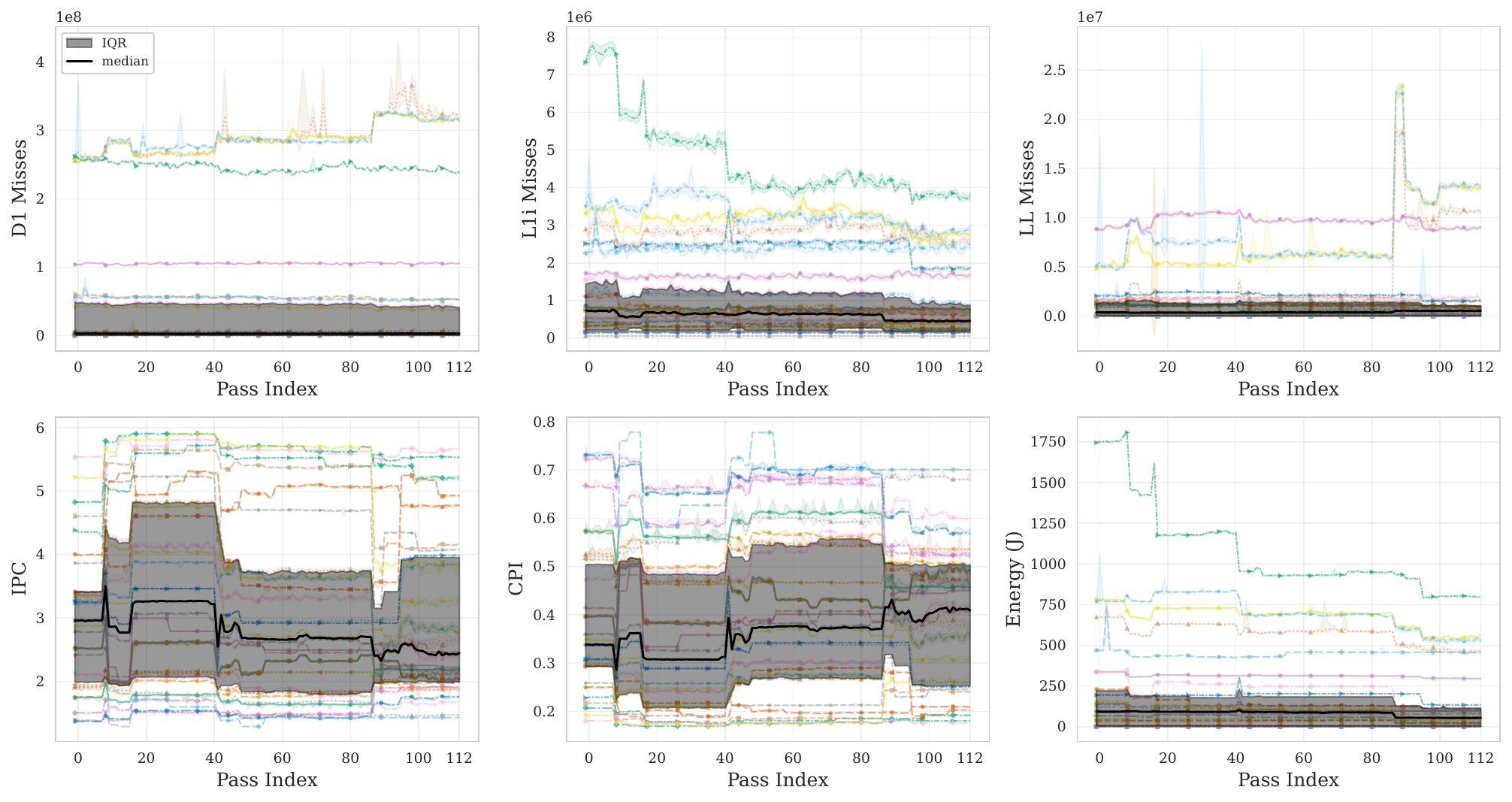}
    \caption{Per-pass evolution of D1, LLC, and L1-I miss rates, IPC, CPI, and energy.}%
    \label{fig:cache-misses}
\end{figure*}

\cref{fig:cache-misses} follows six metrics along the pipeline: each colored line is a kernel, and the grey line and band are the suite-wide median and IQR.
In the cache and energy panels the cross-kernel spread dwarfs the median (the \texttt{-O0} IQR of D1 misses spans \(0.89\)--\(48.9\times10^{6}\)), so the median hugs the floor while the solvers and \texttt{floyd-warshall} float far above.
Each median advances as a plateau broken only at the recurring hot passes (8--9, 16, 41--42, 87, 95, 106).
D1 misses are nearly flat (\(+2.6\%\) end-to-end). L1-I misses fall by 36.8\% in two drops, at passes~9 and~87. LLC misses rise by 35.6\%, with the main step at pass~87.
IPC rises to its peak at pass~8 (\(2.96\to3.50\)), collapses at pass~41, and ends at \(2.44\), a 17.4\% \textit{decrease} that coexists with large speedups because SIMD code executes far fewer, longer-latency instructions.
Energy (RAPL package) falls by 40.9\% (\(92.2\to54.5\)\,J), with the dominant step at pass~87 (\(-35\%\)). To our knowledge this is the first \textit{per-pass} energy profile of the LLVM pipeline.
The binary growth of \S\ref{sec:rq2} therefore does \textit{not} cause instruction-cache pressure: L1-I misses fall with the instruction count, and its only microarchitectural cost is on the data side, where late unrolling inflates LLC misses on the dense solvers.

\begin{rqbox}{Finding 3 (RQ\textsubscript{3})}
Runtime gains come from large reductions in instruction count and cycles, \textit{not} from higher throughput: IPC \textit{falls} (\(-17.4\%\)) as vectorization trades many cheap instructions for fewer, longer-latency SIMD operations, while the D1 miss rate stays essentially flat.
Energy savings track runtime savings (30--60\% suite-wide median): runtime-targeted passes are \textit{de facto} energy-targeted for compute-bound kernels.
Conversely, \texttt{loop-vectorize} and \texttt{loop-unroll} \textit{lower} L1-I misses (fewer instructions fetched); the binary-bloat trade-off of \S\ref{sec:rq2} surfaces instead as \textit{data} working-set expansion, inflating LLC misses on the dense solvers (\texttt{lu}/\texttt{ludcmp}/\texttt{cholesky}).
Branch-misprediction behavior is essentially flat: control-flow optimization is not on the critical path for PolyBench.
\end{rqbox}

\begin{figure}[t]
    \centering
    \includegraphics[width=\columnwidth]{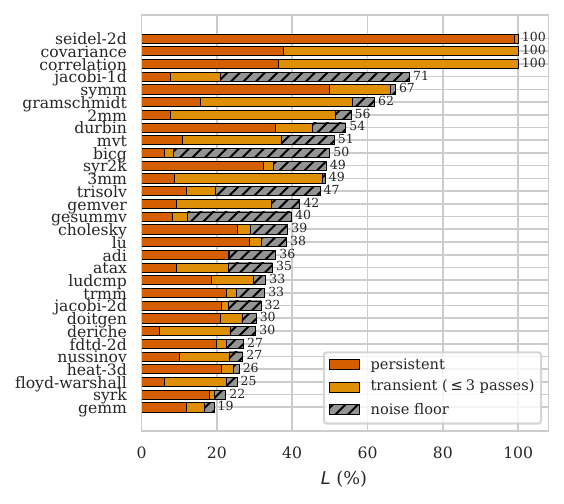}
    \caption{Non-monotonicity loss \(L\) per kernel (bar end), split into persistent loss, transient dips repaired within three passes, and the share due to transitions with byte-identical binaries (noise floor).}%
    \label{fig:phase-interference}
\end{figure}

\subsection{RQ\textsubscript{4}: Non-Monotonicity Loss}\label{sec:rq4}\label{sec:phase-interference}
The regressions of RQ\textsubscript{1} raise the question of how much speedup the pipeline forfeits by moving backwards.
With \(\Delta S_i = S_i - S_{i-1}\) the per-pass change in speedup over \texttt{-O0}, we define the additive ceiling and the loss
\[
S_\mathrm{opt} = \sum_i \max(\Delta S_i, 0),\quad
L = \frac{S_\mathrm{opt} - {[S_\mathrm{actual}]}_0^{S_\mathrm{opt}}}{S_\mathrm{opt}},
\]
where \(S_\mathrm{actual} = S_\mathrm{final} - 1 = \sum_i \Delta S_i\) and \({[x]}_0^{S_\mathrm{opt}} = \min(\max(x,0), S_\mathrm{opt})\).
\(S_\mathrm{opt}\) is the speedup a pipeline would reach if it kept every gain and suffered no regression, so whenever the final speedup is positive \(L\) is simply the total regression mass divided by the total gain mass: a measure of \textit{non-monotonicity}.
\(L\) is not a causal estimate of harm. Passes are non-additive (\S\ref{sec:llvm-pipeline}), so \(S_\mathrm{opt}\) is generally unattainable, and a dip may be the price of a later gain. We therefore treat \(L\) as an upper bound and decompose it to separate those cases.
Across kernels, mean \(L\) is 46.35\% (median 39.3\%, IQR 30.7\%--53.5\%).

\noindent\textit{Decomposition.}\hquad%
\cref{fig:phase-interference} splits each kernel's \(L\) into three parts.
The \textit{noise floor} is the part that disappears when transitions with byte-identical binaries are set to \(\Delta S_i=0\): on average 9.7 points, up to 50 on \texttt{jacobi-1d}.
Of the remaining regression mass, a dip is \textit{transient} to the extent that it is recovered by the gains of the next three passes, and \textit{persistent} otherwise.
Transient dips account for 15.3 points on average (41.9\% of the noise-free loss, 38.6\% when averaged per kernel). The canonical example is \texttt{correlation}, where the drop of pass~41 (\texttt{licm}, \(\Delta S{=}{-}0.26\)) is 71\% undone by pass~42 (\texttt{loop-rotate}, \(\Delta S{=}{+}0.18\)).
Persistent loss averages 21.3 points (median 18.3) and is the component that a pipeline change could, at best, recover.
The four kernels where it is largest, \texttt{seidel-2d}, \texttt{symm}, \texttt{covariance}, and \texttt{correlation}, are exactly those for which RQ\textsubscript{2} found a strictly smaller and faster prefix, so a part of their loss is realizable by stopping early.

\noindent\textit{Attribution.}\hquad%
Persistent loss is concentrated: the first \texttt{instcombine} (pass~16, 23 kernels) contributes 29.9\% of it, \texttt{gvn} (pass~55) 18.5\%, mostly from \texttt{seidel-2d}, and the first \texttt{licm} (pass~41) 8.8\%. By pass type, \texttt{instcombine}, \texttt{gvn}, \texttt{licm}, and \texttt{simplifycfg} account for 78\%.
The passes that deliver most of the speedup are thus also the main sources of the loss that survives to the end of the pipeline.
Unlike prior phase-ordering work, which measures loss relative to an alternative ordering discovered by search~\cite{Kulkarni06,Almagor04,Ashouri17,HajAli20b}, \(L\) is search-free and attributed to single pass positions of the default \texttt{-O3} pipeline.

\begin{rqbox}{Finding 4 (RQ\textsubscript{4})}
The additive upper bound on speedup forfeited by non-monotone behavior is \(L{=}46.35\%\) on average (median 39.3\%), but most of it is not recoverable loss: 9.7 points are measurement noise and 15.3 points are transient dips repaired within three passes.
Persistent loss averages 21.3 points, is largest on four kernels (\texttt{seidel-2d}, \texttt{symm}, \texttt{covariance}, \texttt{correlation}), and is caused mainly by \texttt{instcombine}, \texttt{gvn}, and \texttt{licm}, the same canonicalization core that drives the gains.
\end{rqbox}

\section{Discussion \& Implications}\label{sec:discussion}
\noindent\textbf{Pipeline Surgery.}\quad\label{sec:pipeline-surgery}
The data suggest four concrete uses, each with its limits.
\begin{compactenum}
   \item \textit{Truncating \texttt{-O3} is not a general win.} The faster prefixes of RQ\textsubscript{2} sit in the last 15 passes and gain under 1\% in 20 of 30 kernels. Only for the four kernels with large persistent loss does a per-program ``stop early'' knob pay off (2.5--6.5\%), which makes it a candidate autotuning dimension, not a default.
   \item \textit{Phase-ordering search can be focused.} Persistent loss is concentrated in three positions (the first \texttt{instcombine}, \texttt{gvn}, the first \texttt{licm}). A repair search that perturbs only these positions, or guards them with ``do not run \(p_i\) on IR matching signature \(X\)'', explores a space orders of magnitude smaller than full reordering.
   \item \textit{Inert passes are a free pruning signal.} Here, 58 of 112 positions never modify the IR of any kernel, and 71.8\% of prefixes yield a binary identical to their predecessor. An autotuner can hash IR or binaries to skip redundant measurements and to drop never-firing positions from a program's search space. A position inert in the canonical order may still enable later passes or matter in a reordered pipeline, so pruning decisions must be revalidated end-to-end.
   \item \textit{The canonicalization core is load-bearing in both directions.} The unsupervised ranking recovers the hand-tuned core (\texttt{early-cse}, \texttt{licm}, \texttt{instcombine}, \texttt{loop-vectorize}, \texttt{loop-rotate}, \ldots), and the same core causes most of the persistent loss. Tuning its heuristics (e.g., the iteration limit of the first \texttt{instcombine}) is therefore more promising than pruning the long tail.
\end{compactenum}

\noindent\textbf{Which Programs Benefit.}\quad
Pass effects are strongly program-specific. Universal passes (\texttt{early-cse}, \texttt{sroa}, \texttt{loop-rotate}, \texttt{loop-unroll}: 23--24 kernels with a significant change) coexist with selective ones (\texttt{gvn}: 7 kernels).
Dense linear algebra and stencils profit most from \texttt{loop-vectorize} and \texttt{loop-unroll}, the triangular solvers pay for the same passes in LLC misses, and the \texttt{datamining} reductions end near \(1.0\times\) because mid-pipeline gains are undone later.
A per-program pipeline should therefore key on loop structure and working-set size, which are cheap static features, not on IR size, which RQ\textsubscript{2} shows to be uninformative.

\noindent\textbf{Implications for Cost Models and MLIR.}\quad
IR-instruction count and runtime are essentially uncorrelated both across and within kernels (\cref{fig:ir-vs-runtime}).
Any cost model that uses IR count as a dominant runtime proxy will misrank pipeline variants and must be augmented with dynamic or microarchitectural signals (cycles, IPC, D1 misses)~\cite{Ashouri18, Cavazos07}.
Conversely, counters such as retired instructions and cycles are near-deterministic, so Bayesian-optimization priors built over them can treat per-pass effects as essentially deterministic.
For compute-bound kernels, passes that reduce runtime also reduce energy by 30--60\% (\cref{fig:cache-misses}), so \texttt{-O3} is a free energy intervention for this class. Transfer to memory-bound, batched, or accelerator-hosted ML requires validation (\S\ref{sec:threats}).
On the dense solvers, late \texttt{loop-vectorize} and \texttt{loop-unroll} expand the data working set and inflate LLC misses. An MLIR lowering pipeline could gate these passes behind a working-set/LLC-capacity predicate.

\noindent\textbf{Methodology Recommendations.}\quad
Pass-level studies should calibrate significance on a null. In our data, \(p<0.05\) alone flags 25.7\% of transitions whose binaries are byte-identical, so naive per-pass testing would report hundreds of spurious effects. Combining an effect-size floor with a test, and checking binary identity, reduces this to 0.9\%.
Reporting mean and median jointly, and absolute and relative impact jointly, caught estimator- and scale-driven artifacts at negligible cost.
All recommendations are conditioned on PolyBench (default dataset), LLVM 21.1.8, \texttt{-O3}, single-threaded execution, and Intel Alder Lake. Re-validation on SPEC CPU, \texttt{-Os}/\texttt{-Oz}, and other platforms is required before generalizing.

\section{Threats to Validity}\label{sec:threats}
The discussion follows the taxonomy of Wohlin \emph{et al.}~\cite{Wohlin12}.

\noindent\textbf{Conclusion Validity.}\hquad%
We keep the 95\% confidence interval under 1\% of the mean for every metric and report median and mean jointly~\cite{Georges07, Mytkowicz09}, so estimator-driven artifacts surface instead of hiding.
Every significance claim uses a criterion calibrated on 2,434 byte-identical transitions, with an empirical false-positive rate of 0.9\%. Benjamini--Hochberg correction across the \(113\times30\) grid changes the counts by at most one.
We dropped Cohen's \(d\) because vanishing variance on near-deterministic transitions inflates it without bound.
The decomposition of \(L\) depends on the recovery window. With one pass instead of three, mean persistent loss is 24.2 points, and with ten it is 16.4, so the qualitative conclusion (most of \(L\) is not persistent) does not hinge on this choice.
Finally, \(L\) is an upper bound on forfeited speedup, not an estimate of what is recoverable.

\noindent\textbf{Internal Validity.}\hquad%
The differential design ascribes the \(B_{i-1}\!\to\!B_i\) delta to \(p_i\), but this delta is context- and order-dependent: a pass inert in the canonical \texttt{-O3} order may be load-bearing for later passes or in a reordered pipeline.
We thus refrain from inferring global dispensability and require end-to-end revalidation before pruning (\S\ref{sec:pipeline-surgery}).
System confounds are controlled as in \S\ref{sec:methodology} (core pinning, RT-FIFO scheduling, \texttt{performance} governor, disabled ASLR), with cache references, RSS, page faults, and context switches as checks. Runtime and counters are collected separately to avoid instrumentation overhead.
The residual drift visible on byte-identical binaries is absorbed by the calibrated criterion.

\noindent\textbf{Construct Validity.}\hquad%
``Marginal pass impact'' is operationalized as the in-context prefix delta, which measures contribution \textit{within} the canonical order. This matches how \texttt{-O3} deploys passes but is not a standalone-effect measurement.
The pass sequence is extracted statically (\texttt{opt --print-pipeline-passes}) and flattened. Each prefix keeps the pass-manager nesting, so a function pass runs once per function and loop passes run per loop, exactly as in \texttt{-O3}. On-demand analyses, however, are not enumerated, and a pass that is invoked on several functions counts as a single transition.
We complement the static view with the dynamic evidence that matters for attribution, namely whether the IR and binary actually change at each transition (\S\ref{sec:rq1}).
Runtime, \texttt{clang+opt+llc} compile time, and binary size are direct measures. RAPL package energy excludes the DRAM and GPU domains, under-capturing full-system energy for memory-bound workloads.

\noindent\textbf{External Validity.}\hquad%
The findings are conditioned on PolyBench/C 4.2.1 (default dataset), LLVM 21.1.8, \texttt{-O3}, and single-threaded execution on one Intel Alder Lake host.
PolyBench's regular affine loop nests isolate single-pass effects but do not represent irregular, pointer-heavy, or multi-threaded code. Its default working sets fit in the LLC, so memory-bound behavior is under-represented.
Larger suites would mask single-pass effects or exceed our repetition budget (\S\ref{sec:methodology}), so cross-suite, larger-dataset, and cross-architecture replication is required before generalizing.
We expect structural findings (concentration in the canonicalization core, non-monotonicity, back-loading) to transfer best, because they reflect the design of the pipeline, while energy and LLC/working-set effects depend on the host's RAPL domain and cache hierarchy.
The ranking is version-locked. The MLIR transfer argument rests on structural similarity (\S\ref{sec:methodology}) and is an unvalidated hypothesis.

\section{Related Work}\label{sec:related}
We are not aware of any systematic study quantifying individual LLVM pass impact across multiple performance dimensions. We review the relevant strands below.

\noindent\textbf{Search-Based and Learning-Based Optimization.}\hquad%
Iterative compilation~\cite{Bodin98, Cooper99, Cooper02} explores the optimization space by evaluating compiler configurations: small search budgets can outperform \texttt{-O3}~\cite{Almagor04, Cooper99} with program-dependent optimal configurations~\cite{Kulkarni03, Purini13, Eeckhout03}, while heuristic~\cite{Pan06, Pan08, Kulkarni04b} and ML~\cite{Agakov06, Kulkarni12, Fursin11} approaches, collaborative filtering~\cite{Liu21}, and graph-based models~\cite{Nobre16} address the combinatorial space.
Autotuning~\cite{Ashouri18} adds Bayesian optimization~\cite{Ashouri16}, neural predictors~\cite{Ashouri17, Kulkarni12}, and reinforcement learning (AutoPhase~\cite{HajAli20b}, CompilerGym~\cite{Cummins22}), with recent work reducing search cost via multi-phase learning~\cite{Zhu24}, sub-sequence modeling~\cite{Pan25}, and proxy metrics~\cite{Zhao25}; a recurring limitation is reliance on expensive or inaccurate performance signals.
More broadly, ML links hardware counters to optimization decisions~\cite{Cavazos07} or uses deep learning for vectorization~\cite{HajAli20} and phase ordering~\cite{HajAli20b}, yet surveys~\cite{Ashouri18} highlight the lack of fine-grained training data and difficulty isolating pass-level effects.
Our work is orthogonal and complementary: we quantify each pass's marginal contribution across runtime, binary size, and counters, producing fine-grained data for supervision, surrogate targets, or priors while filling this pass-level data gap.

\noindent\textbf{Compiler Fuzzing and Testing.}\hquad%
Csmith~\cite{Yang11}, EMI~\cite{Le14}, YARPGen~\cite{Livinskii20}, and CLsmith~\cite{Lidbury15} target correctness, not performance.
Our work is complementary: correctness enables reliable measurement, while performance characterization can guide fuzzing toward less understood optimizations.

\noindent\textbf{Closest Work.}\hquad%
Three lines of work are closest to ours, and they differ in objective, method, and findings.
De La Torre \textit{et al.}~\cite{DeLaTorre18} share our objective of attributing performance to LLVM passes, but they study a limited set of passes and execution time only. We cover the complete \texttt{-O3} pipeline in its deployed order, add four further dimensions, and calibrate significance against a null, which lets us separate real regressions from noise.
Chen \textit{et al.}~\cite{Chen12} deconstruct iterative optimization in GCC/ICC and find that a few optimizations carry most of the benefit. Our concentration result agrees, on a different compiler family, under scale-free measures, and on four further dimensions, which suggests that skewed pass impact is a property of optimization pipelines and not of a specific compiler.
Phase-ordering studies~\cite{Kulkarni06, Jantz14, Almagor04} measure what a better order could gain, relative to orders found by search. We instead ask how much the \textit{default} order forfeits along its own trajectory, which needs no search, and find that most of the apparent loss is noise or transient.

\noindent\textbf{Performance Analysis of Compiler Optimizations.}\hquad%
Cavazos \textit{et al.}~\cite{Cavazos07} use counters to predict optimization benefits, Demertzi \textit{et al.}~\cite{Demertzi11} analyze reliability, Popescu and Lopes~\cite{Popescu25} study undefined behavior, and Tan \textit{et al.}~\cite{Tan20} identify missed optimizations.
None provides a cumulative, multi-dimensional evaluation of the full LLVM \texttt{-O3} pipeline.

\noindent\textbf{Superoptimization and PGO.}\hquad%
Our ranking, dominated by \texttt{instcombine} and \texttt{early-cse}, corroborates superoptimization results~\cite{Massalin87, Bansal06}: peephole algebraic simplifications account for a disproportionate share of achievable optimizations.
Our finding that pass-budget decisions can be made per pipeline concerns \textit{compile-time allocation}, whereas PGO/FDO~\cite{Chen16b} shows that \textit{which transformations to apply} benefits from per-program profiles. The two are compatible.

\section{Conclusion}\label{sec:conclusion}
We presented a systematic, multi-dimensional study of the LLVM \texttt{-O3} pipeline, decomposing it into cumulative per-pass prefixes across \(84{,}750\) measurements over 113 variants of 30 PolyBench/C kernels, and calibrating every significance claim on the transitions that leave the binary unchanged.
The pipeline is non-monotone (36\% of significant runtime changes are regressions) and back-loaded, and a small core of passes drives the gains under both absolute and scale-free measures.
The final \texttt{-O3} binary is only marginally Pareto-dominated, except on four kernels where a strictly smaller and faster prefix exists.
IR-instruction count is an unreliable runtime proxy, and runtime-targeted passes are \textit{de facto} energy-targeted (30--60\%).
The search-free upper bound on speedup forfeited by non-monotonicity is \(L=46.35\%\), but only 21.3 points of it persist to the end of the pipeline, mostly caused by the same canonicalization passes that deliver the speedup.
These results point pass pruning, phase-ordering repair, and cost-model calibration to a handful of pass positions.
Future work targets SPEC CPU, larger datasets, \texttt{-Os}/\texttt{-Oz}, more platforms, and MLIR-lowered ML kernels.


\bibliographystyle{IEEEtran}
\bibliography{local,strings,metrics,compilers,data_structures,programming,software_engineering,software_architecture,dsl,pl,splc,oolanguages,my_work,grammars,ml+nn,security,roles,learning,cop,testing,dsu,distributed_systems,reflection,aosd,foundations,petri-nets,pattern,logic}

\end{document}